\documentclass{SciPost}

\hypersetup{
    colorlinks,
    linkcolor={red!50!black},
    citecolor={blue!50!black},
    urlcolor={blue!80!black}
}

\usepackage[bitstream-charter]{mathdesign}
\DeclareSymbolFont{usualmathcal}{OMS}{cmsy}{m}{n}
\DeclareSymbolFontAlphabet{\mathcal}{usualmathcal}

\fancypagestyle{SPstyle}{
\fancyhf{}
\lhead{\colorbox{scipostblue}{\bf \color{white} ~SciPost Physics Proceedings }}
\rhead{{\bf \color{scipostdeepblue} ~Submission }}

\fancyfoot[C]{\textbf{\thepage}}
}

\usepackage[utf8]{inputenc}      % input font encoding

\usepackage{amsmath,mathtools}
\usepackage[T1]{fontenc}          % output font encoding
\usepackage{booktabs,tabularx}
\usepackage{graphicx}
\usepackage{xspace}
\usepackage{lmodern}
\usepackage{geometry}
\usepackage{todonotes}
\usepackage{listings}
\usepackage[absolute]{textpos}
\usepackage[many]{tcolorbox}
\usepackage{xparse}
\usepackage[font=small,labelfont=bf,format=plain,margin=0.05\textwidth]{caption}
\usepackage{bbm}
\usepackage{tabularx}
\usepackage{soul}
\usepackage[capitalize]{cleveref}
\usepackage{slashed}
\usepackage{pifont}
\usepackage[shortlabels]{enumitem}
\usepackage{nicefrac}
\usepackage{subfig}
\usepackage[super]{nth}
\usepackage{lineno}
\usepackage{authblk}

\newcommand{\cp}{\ensuremath{{\cal CP}}\xspace}

\newcommand{\ttH}{\ensuremath{t\bar t H}\xspace}
\newcommand{\ttbar}{\ensuremath{t\bar t}\xspace}

\newcommand{\pTx}[1]{\ensuremath{p_{T,{#1}}}\xspace}
\newcommand{\tstar}{\ensuremath{\theta^*}\xspace}

\newcommand{\tev}{\,\, \mathrm{TeV}}

\newcommand{\invfb}{\,\, \mathrm{fb}^{-1}}

\definecolor{Darkgreen}{rgb}{0.,.7,0.2}
\definecolor{Darkblue}{rgb}{0.,.2,0.7}
\definecolor{Magenta}{rgb}{0.7,0.,0.7}

\newcommand{\gt}{\ensuremath{g_t}\xspace}
\newcommand{\at}{\ensuremath{\alpha_t}\xspace}

\def\be{\begin{equation}}
\def\ee{\end{equation}}

\begin{document}

\pagestyle{SPstyle}

\begin{center}{\Large \textbf{\color{scipostdeepblue}{
%%%%%%%%%% TODO: Write your article's title here
Proposal for simplified template cross-sections extension using \cp observables in \ttH \\
%%%%%%%%%% END TODO: TITLE
}}}\end{center}

\begin{center}\textbf{
%%%%%%%%%% TODO: AUTHORS
% Write the author list here. 
% Use (full) first name (+ middle name initials) + surname format.
% Separate subsequent authors by a comma, omit comma and use "and" for the last author.
% Mark the corresponding author(s) with a superscript symbol in this order
% \star, \dagger, \ddagger, \circ, \S, \P, \parallel, ...
Alberto Carnelli\textsuperscript{1$\dagger$},
%%%%%%%%%% END TODO: AUTHORS
}\end{center}

\begin{center}
%%%%%%%%%% TODO: AFFILIATIONS
% Write all affiliations here.
% Format: institute, city, country
{\bf 1} LAPP, Université Savoie Mont Blanc, CNRS/IN2P3, 74941 Annecy; France
\\

%%%%%%%%%% END TODO: AFFILIATIONS
%%%%%%%%%% TODO: EMAIL
% Provide email address of corresponding author(s)
%\\[\baselineskip]
$\dagger$ \href{mailto:alberto.carnelli@cea.fr}{\small alberto.carnelli@cern.ch}
%%%%%%%%%% END TODO: EMAIL
\end{center}

\definecolor{palegray}{gray}{0.95}
\begin{center}
\colorbox{palegray}{
  \begin{tabular}{rr}
  \begin{minipage}{0.36\textwidth}
    \includegraphics[width=60mm,height=1.5cm]{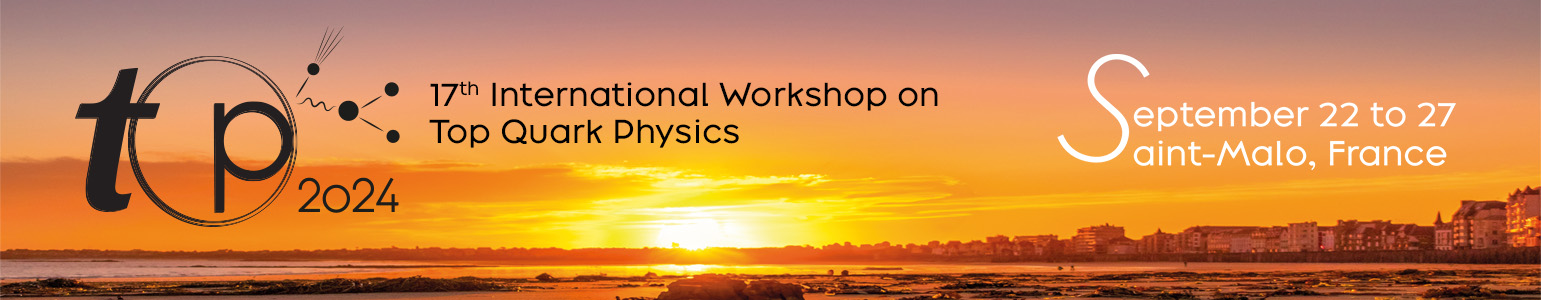}
  \end{minipage}
  &
  \begin{minipage}{0.55\textwidth}
    \begin{center} \hspace{5pt}
    {\it The 17th International Workshop on\\ Top Quark Physics (TOP2024)} \\
    {\it Saint-Malo, France, 22-27 September 2024
    }
    \doi{10.21468/SciPostPhysProc.?}\\
    \end{center}
  \end{minipage}
\end{tabular}
}
\end{center}

\section*{\color{scipostdeepblue}{Abstract}}
\textbf{\boldmath{%
%%%%%%%%%% TODO: ABSTRACT
% Write your abstract here.
The Large Hadron Collider (LHC) offers a unique opportunity to investigate \cp\ violation in the Yukawa coupling between the Higgs boson and the top quark by studying Higgs production in association with top quarks; this is of fundamental importance, seeing that the \cp\ properties of the Higgs boson are yet to measure with high precision. To address this, the focus of this work has been an extension of the simplified template cross-section (STXS) framework, devised to be sensitive to \cp\ effects. Our study focused on \cp-sensitive observables across multiple Higgs decay channels, comparing their performances. The result indicates that the most efficient extension of the current binning used in the STXS framework, which currently uses the Higgs boson's transverse momentum \pTx{H}, requires adding one further split using \cp-sensitive observables. Between these observables, one of the best is the Collins-Soper angle $|\cos\theta^*|$, a variable derived from momenta information of the top quarks. We have investigated the improvement brought by our two-dimensional STXS setup and compared it to the currently employed methodologies, finding an increase in performances at an integrated luminosity of $300\invfb$. Moreover, our results highlight that this advantage seems to be present also at $3000\invfb$.
%%%%%%%%%% END TODO: ABSTRACT
}}

\vspace{\baselineskip}

%%%%%%%%%% BLOCK: Copyright information
% This block will be filled during the proof stage, and finilized just before publication.
% It exists here only as a placeholder, and should not be modified by authors.
\noindent\textcolor{white!90!black}{%
\fbox{\parbox{0.975\linewidth}{%
\textcolor{white!40!black}{\begin{tabular}{lr}%
  \begin{minipage}{0.6\textwidth}%
    {\small Copyright attribution to authors. \newline
    This work is a submission to SciPost Phys. Proc. \newline
    License information to appear upon publication. \newline
    Publication information to appear upon publication.}
  \end{minipage} & \begin{minipage}{0.4\textwidth}
    {\small Received Date \newline Accepted Date \newline Published Date}%
  \end{minipage}
\end{tabular}}
}}
}
%%%%%%%%%% BLOCK: Copyright information

%%%%%%%%%% TODO: LINENO
% For convenience during refereeing we turn on line numbers:

% You should run LaTeX twice in order for the line numbers to appear.
%%%%%%%%%% END TODO: LINENO

%%%%%%%%%% TODO: TOC 
% Guideline: if your paper is longer that 6 pages, include a TOC
% To remove the TOC, simply cut the following block
%\vspace{10pt}
%\noindent\rule{\textwidth}{1pt}
%\tableofcontents
%\noindent\rule{\textwidth}{1pt}
%\vspace{10pt}
%%%%%%%%%% END TODO: TOC

%%%%%%%%% TODO: CONTENTS 

% Write your article contents here, starting from first \section.
% An example structure is given below.

\section{Introduction}
\label{sec:intro}

Current observations indicate a baryon asymmetry in our Universe~\cite{Gavela:1993ts,Huet:1994jb}, an asymmetry that can not be described by the Standard Model (SM) of particle physics. New sources of \cp violation not expected by the current SM theory are then needed and the search for corresponding \cp-violating interactions is an essential target for searches beyond the SM (BSM) at the LHC.
Recently, the \cp structure of the Higgs--fermion interactions has started to be probed ~\cite{CMS:2020cga,ATLAS:2020ior}. It has to be noted that BSM theories allow for a larger amount of \cp violation in the Yukawa couplings with respect to other possible sources, like, for example, coming from interactions with massive vector bosons that are loop-suppressed. It is then of high importance to focus on identifying the \cp situation of Higgs--fermion interactions, and of these, the top-Yukawa coupling has a special relevance being the highest. The Higgs Characterization Model~\cite{Artoisenet:2013puc} allows us to describe possible \cp violation by varying the SM interaction as follows:

%While the \cp structure of the Higgs couplings to massive gauge bosons is already relatively tightly constrained~\cite{ATLAS:2015zhl,ATLAS:2016ifi,CMS:2017len,CMS:2019jdw,CMS:2019ekd,ATLAS:2020evk,ATLAS:2021pkb,CMS:2022mlq,ATLAS:2022tan,ATLAS:2023mqy}.
%, the test of the \cp structure of the Higgs--fermion interactions at the LHC only started recently~\cite{CMS:2020cga,ATLAS:2020ior,CMS:2021sdq,CMS:2021nnc,CMS:2022dbt,CMS:2022mlq,ATLAS:2022akr,ATLAS:2023cbt}. Moreover, most BSM theories predict a larger amount of \cp violation in the Yukawa couplings than in the interactions with massive vector bosons, since the latter are loop-suppressed.
%

\begin{align}
    \mathcal{L}_\text{top-Yuk} = \frac{y_t^\text{SM}\gt}{\sqrt{2}}\bar t\left(\cos\at + i \gamma_5 \sin\at\right)t H\,.
\label{eq:top-Yukawa}
\end{align}
In this parametrization:
\begin{itemize}
\item $y_t^\text{SM}$ is the SM top-Yukawa coupling,
\item $g_t$ acts as modifier of the strength of the top-Yukawa coupling,
\item $\alpha_t$ is the \cp-mixing angle.
\end{itemize}
We retrieve the SM here if we consider $\gt=1$ and $\at=0$.
This model has been used for direct searches in different Higgs decay channels. These searches have brought some initial limits but usually have used a strategy focusing on the specific Higgs decay channels under study, its detector environments, and with the channels specific background assumptions. These analyses also have been constructed on particular signal models, so they became dependent on assumptions of other involved couplings done in these models. %tailored to the various channels and are not easy to combine. %typically tailored to specific Higgs decay channels, detector environments, and background assumptions. Moreover, they are designed for specific signal models and intrinsically depend on the assumptions about the other couplings involved. %As a consequence, the combination of different decay channels (see e.g.~\cite{CMS:2021nnc,CMS:2022dbt}) and across experiments is non-trivial. The usage of such approaches also makes the reinterpretation of the analyses difficult, as it usually requires the signal efficiency as a function of the final discriminating algorithm output score. 
The simplified template cross-section (STXS) framework~\cite{Amoroso:2020lgh} has been explicitly established to mitigate issues like channel-specific setup. The objective of this work is, using \cp observables, to enhance the sensitivity in \cp of the STXS framework for the \ttH production by adding an additional variable to the Higgs boson's transverse momentum \pTx{H} currently employed.

\section{Study setup}
\label{sec:obs}
Using \texttt{MadGraph5\_aMC@NLO}~\cite{Alwall:2014hca} (version 3.3.2) we produced parton-level events for the $pp \rightarrow t\bar{t}H$ process at a center-of-mass energy of $\sqrt{s}=13\tev$. All samples were produced at leading order (LO) with one million events each, applying a scaling factor of 1.14~\cite{Demartin:2014fia} as a next-to-leading-order (NLO) correction. The effect of NLO correction to the \ttH total cross-section, as explained in~\cite{Demartin:2014fia,Bahl:2020wee}, is only weakly dependent on the \cp part of the top-Yukawa coupling. This observation is still valid when considering the various kinematic distributions in this channel. We evaluate 11 \cp-discriminating observables across four reference frames:

\begin{itemize}
    \item the laboratory frame (lab frame),
    \item the $t\bar t$ rest frame, where $\boldsymbol{p}_t+\boldsymbol{p}_{\bar t}= \boldsymbol{0}$ ($t\bar t$ frame),
    \item the $H$ rest frame, where $\boldsymbol{p}_H=\boldsymbol{0}$ ($H$ frame),
    \item the $t\bar tH$ rest frame, where $\boldsymbol{p}_t+\boldsymbol{p}_{\bar t}+\boldsymbol{p}_H= \boldsymbol{0}$ ($t\bar tH$ frame).
\end{itemize}

%The \cp-sensitive observables considered in this work are summarised in the appendix in Table~\cref{tab:obs}. We analyse them in all applicable rest frames listed above. %Any new observable candidate for the extended STXS framework should be defined for all \ttH\ events, regardless of the Higgs boson and top quark pair decay mode. Therefore, in the following, we consider that the Higgs boson, the top quark and anti-top quark are reconstructed experimentally such that their momenta are accessible. In practice, we compute the observables at the parton level and apply specific acceptance and smearing factors to mimic detector effects.

\section{Performance evaluation}
\label{sec:another}

To extend the STXS framework, we focus on the three channels studied in current $t\bar{t}H$ analyses at the LHC that target different Higgs final states: $t\bar{t}H(\to\gamma\gamma)$, $t\bar{t}H(\to b\bar{b})$, and $t\bar{t}H$(multilep.), which is a way to group together $H(\to\tau\tau)$, $H(\to W^{+}W^{-})$ and $H(\to ZZ)$ decays considering multiple leptons in the final state. We take the assumption to have access to measurements of the $t\bar{t}H$ distributions in each channel in order to measure the sensitivity of the observables discussed in~\cref{sec:obs} to \cp violation. The effect of reconstruction effects (detector performance, etc) are taken into account by using scaling and smearing factors to the parton-level $t\bar{t}H$ events to produce realistic yields. These effects are obtained from the most recent results of the ATLAS and CMS collaborations in each of the studied channels and can be consulted in~\cite{ATLAS:2020ior,ATLAS:2023cbt,CMS:2020cga,CMS:2021nnc,CMS:2022mlq,CMS:2022dbt}.  

The sensitivity of an observable to a given BSM model is quantified through a significance $S$, as utilized by the ATLAS collaboration~\cite{ATL-PHYS-PUB-2020-025}. This significance reflects the power to distinguish the BSM hypothesis, parameterized by $g_t$ and $\alpha_t$, from the SM hypothesis, which implies $g_t=1$ and $\alpha_t=0$.

\section{Results}

The result presented here considers the whole dataset expected to be available at the end of LHC Run-3, corresponding to a luminosity of $300\invfb$. The current experimental limits exclude $g_t=1$ and $\at \gtrsim 43^\circ$ at the $95\%~\mathrm{CL}$ with $139\invfb$~\cite{ATLAS:2020ior}, as consequence a benchmark of $g_t=1$ and $\at = 35^\circ$ was chosen. We calculate the significance for all the studied observables and the associated two-dimensional observable combinations using 6 bins for every observable. The results indicate that using a combination of two observables gives higher performances. We favor combinations with \pTx{H} due to the existing STXS binning and seeing that the results are similar to the optimal combination of the other two observables, choosing the binning of a further observable, ensuring that each bin remains populated by at least a few events. The best observables to combine with \pTx{H} are: $\Delta\phi_{t\bar{t}}^\text{lab}$, $b_1^\text{lab}$, $b_2^\text{lab}$ (\cite{Gunion:1996xu}), $\Delta\eta_{t\bar{t}}^{\ttbar}$ and $|\cos\theta^*|$ (\cite{Collins:1977iv,Goncalves:2018agy}). After further bin optimization and background studies, we further reduced the candidate observables to $b_2^\text{lab}$, $\Delta\eta_{t\bar{t}}^{\ttbar}$, and $|\cos\theta^*|$ with the following binning: 
\begin{itemize}
    \item $b_2^{\mathrm{lab}}$: [-1, -0.6, -0.4, -0.2, 0., 0.3, 1.0],
    \item $\Delta\eta_{t\bar t}^{t\bar t}$: [0, 0.5, 1, 1.5, 2, 3, 5],
    \item $|\cos\theta^{*}|$: [0, 0.2, 0.4, 0.55, 0.7, 0.85, 1].
\end{itemize}

\begin{figure}[!htbp]
    \centering
    \includegraphics[width=.54\textwidth]{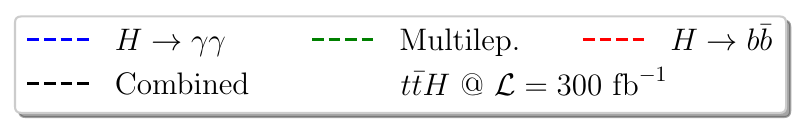} \\
    \includegraphics[trim=1.cm 0.25cm 1.cm 0.25cm, width=.48\textwidth]{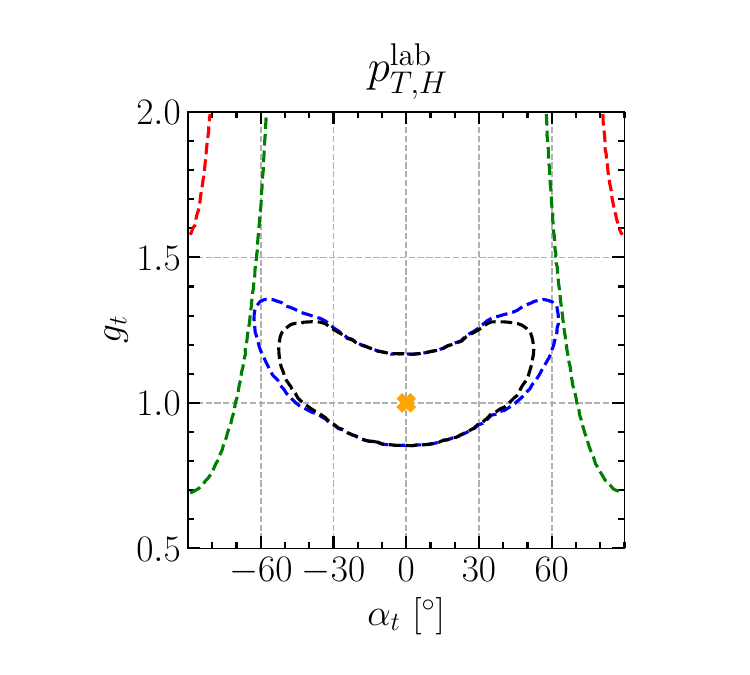}
    \includegraphics[trim=1.cm 0.25cm 1.cm 0.25cm, width=.48\textwidth]{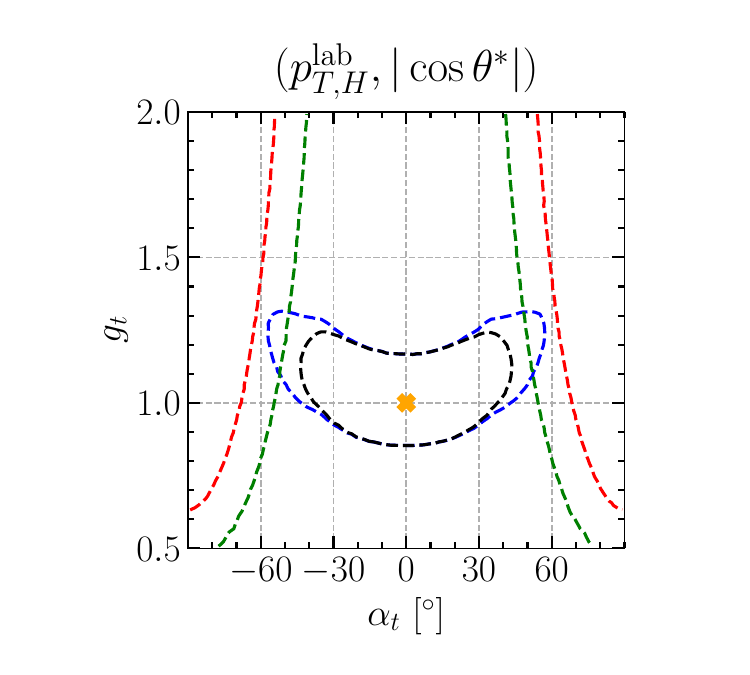}
    \caption{Limits obtained using our samples at the 95\% confidence level in the plane $(\at,\gt)$ expected at the end of LHC Run-3 ${\mathcal{L} = 300\invfb}$ with the current $\pTx{H}$ STXS setup (left) and the improvement obtained with one of our 2-dimensional STXS extension: $(\pTx{H},|\cos\theta^*|)$ (right).}
    \label{fig:limits}
\end{figure}

Based on this, we propose to extend the current STXS binning by adding one of the above observables; an example showing the improvement in exclusion limits adding $|\cos\theta^*|$ is presented in Fig~\ref{fig:limits}. The results show that further splitting the current STXS $\pTx{H}$ using the $|\cos\tstar|$ observable, then there is a 12\% improvement in the combined limit at $\gt=1$, allowing to reach an exclusion limit for $|\at| \lesssim 36^\circ$. In this scenario, we have a maximum $40$\% improvement and is reached at $\gt=1.24$ considering the $\ttH(\to\gamma\gamma)$ standalone limit. We also use a simplistic extrapolation to quantify the impact of this expanded framework on the exclusion limits at 3000 fb$^{-1}$, observing similar improvement. We performed a further study with Boosted Decision Trees (BDT), taking advantage of all the available \cp observables, trying to obtain the maximal discrimination, showing only $\sim 10 \%$ improvement in performances.

\section{Conclusion}

To propose an extension of the current STXS v1.2 binning in $\pTx{H}$, we selected \cp observables and studied their performances and combination; the results highlight that is ideal to extend the current STXS binning that uses the $\pTx{H}$ by combining with: $\Delta\eta_{t\bar{t}}$, $|\cos\theta^*|$, and $b_2$. In the $(g_t, \alpha_t)$ plane, we have evaluated the exclusion limits using our proposed STXS extension versus the current STXS. For both $300\invfb$ and $3000\invfb$ of data, the proposed two-dimensional binning outperforms the current STXS binning and we also evaluate the result that can be obtained by employing a different methodology, a BDT, observing similar performances.

\begin{appendix}

\bibliography{SciPostPRTOP2024CarnelliPoster_final.bib}

%at the end of your document. If you are not using our LaTeX template, please still use our bibstyle %as
%\begin{verbatim}
%\bibliographystyle{SciPost_bibstyle}
%\end{verbatim}
%in order to simplify the production of your paper.
\end{appendix}

%%%%%%%%% END TODO: CONTENTS

%%%%%%%%%% TODO: BIBLIOGRAPHY
% Provide your bibliography here. You have two options:

%%% FIRST OPTION
% Write your entries here directly, following the example below, including:
% Author(s), Title, Journal Ref. with year in parentheses at the end, followed by the DOI number.

% \begin{thebibliography}{99}
% \bibitem{1931_Bethe_ZP_71}
% H. A. Bethe, \textit{Zur Theorie der Metalle. i. Eigenwerte und Eigenfunktionen der linearen Atomkette}, Zeit. f{\"u}r Phys. \textbf{71}, 205 (1931), \doi{10.1007\%2FBF01341708}.

% \bibitem{arXiv:1108.2700}
% P. Ginsparg, \textit{It was twenty years ago today...}, (arXiv preprint) \doi{10.48550/arXiv.1108.2700}.

% \end{thebibliography}

%%% SECOND OPTION
% Use your bibtex library, formatted by the SciPost style file.
%\bibliography{SciPost_Example_BiBTeX_File.bib}

%%%%%%%%%% END TODO: BIBLIOGRAPHY

\end{document}